\documentstyle[eqsecnum,aps]{revtex}
\input epsf
\begin{document}

\twocolumn[\hsize\textwidth\columnwidth\hsize\csname
@twocolumnfalse\endcsname

\title{Muonium as a hydrogen analogue in silicon and germanium;
quantum effects and hyperfine parameters}

\author{A.~R.~Porter, M.~D.~Towler and R.~J.~Needs}

\address{TCM Group, Cavendish Laboratory,\\ Madingley Road,
Cambridge, CB3 0HE, UK}

\date{\today}

\maketitle

\begin{abstract} 
\begin{quote}
\parbox{16 cm}{\small 

We report a first--principles theoretical study of hyperfine interactions,
zero--point effects and defect energetics of muonium and hydrogen impurities in silicon and germanium. The spin--polarized density functional method
is used, with the crystalline orbitals expanded in all--electron Gaussian basis
sets. The behaviour of hydrogen and muonium impurities at both the tetrahedral
and bond--centred sites is investigated within a supercell approximation. To
describe the zero--point motion of the impurities, a double adiabatic
approximation is employed in which the electron, muon/proton and host lattice
degrees of freedom are decoupled. Within this approximation the relaxation of
the atoms of the host lattice may differ for the muon and proton, although in
practice the difference is found to be slight. With the inclusion of
zero--point motion the tetrahedral site is energetically preferred over the
bond--centred site in both silicon and germanium. The hyperfine and
superhyperfine parameters, calculated as averages over the motion of the muon,
agree reasonably well with the available data from muon spin resonance
experiments.}
\end{quote}

\end{abstract}

]
\narrowtext

\section{Introduction}

Hydrogen is known to have a wide range of physical effects in semiconductors,
including the passivation of states associated with deep--level impurities,
enhancement of the diffusivity of oxygen, and the formation of large, planar
structures known as platelets.~\cite{Pat88} It is present in large quantities
during the processing stages of device manufacture and is one of the commonest
impurities in technologically important materials such as silicon and
germanium. Since such hydrogen impurities can have significant effects on
semiconductor electrical properties, a more complete understanding of their
behaviour at the microscopic level is desirable.

Paramagnetic hydrogen centres can in principle be studied using the electron
paramagnetic resonance (EPR) technique. Information on their local environment
is obtained by following the time evolution of the signal corresponding to the
coupling of the spin of the impurity with an external electromagnetic field.
However, few studies have been reported for hydrogen in semiconductors because
the hydrogen atoms are mobile and diffuse to defects where they form passivated
complexes. The transient centres of isolated hydrogen impurities are
nevertheless of significant interest because of their involvement in diffusion
processes, and in fact they may be studied using muon spin resonance ($\mu$SR)
techniques. Muons have the same charge as protons but only about one ninth of
the mass. They can capture an electron to form a hydrogen--like bound state
known as muonium (given the symbol Mu), and it is thus possible to
consider the muon as a proton analogue. Transient centres of implanted positive
muons in semiconductors may be studied as the muon has a lifetime of just 2.2
$\mu$s and diffuses to locally stable sites within a few nanoseconds.  The
short lifetime also means that there is almost never more than one muon in the
sample at any one time, and that the distribution of muons does not reach true
thermal equilibrium. In a muon spin resonance  experiment fully polarized
positive muons are injected into the sample, and by observing the positrons
produced by the decay of the muons one can obtain information about the
defect.~\cite{Pat88}

When $\mu$SR experiments are performed on silicon or germanium, two different
hyperfine signals are observed.  One of these is entirely isotropic while the
other has an anisotropic (dipolar) component with uniaxial symmetry along the
[1 1 1] axis.~\cite{Pat88} The impurity responsible for the former signal is
usually referred to as normal muonium (Mu) and that for the latter, anomalous
muonium (Mu$^{*}$).  Normal muonium has been identified as muonium in the
interstitial region, probably in the vicinity of the tetrahedral (T)
interstitial site.  Symons and Cox~\cite{symons} first suggested that anomolous
muonium corresponds to a neutral muonium at the bond--centred (BC) site and
this has been borne out by a number of theoretical studies.  The various
experimental data for muonium in silicon have been interpreted in terms of a
configuration--coordinate diagram.~\cite{config}

The majority of recent theoretical work in this area has been at the first
principles level within an adiabatic approximation, using the local spin
density (LSDA) or generalized gradient (GGA) approximations to density
functional theory (DFT).~\cite{parr&yang} Calculations using pseudopotentials
and plane--wave basis sets~\cite{VdW90,VdW_Blochl93,Luch97} have been
reasonably successful in reproducing the hyperfine and superhyperfine
parameters observed in experiments.  The majority of such calculations appear
to demonstrate that hydrogen impurities at the T and BC sites have similar
energies.~\cite{Luch97,chang89}

The application of the Feynman path--integral~\cite{Feynman} method to the
study of these systems enables the effect of the quantum nature of the muon to
be studied directly at finite temperatures. However, the large computational
demands of such an approach have limited its use to date.  Ram\'{\i}rez and
Herrero~\cite{Ramirez} used the path integral molecular dynamics method to
study hydrogen and muonium in silicon with the H/Mu-Si interaction described by
an empirical three--body potential. However, the results appear to be in
conflict with experiment. Recently, Miyake {\it et al.}~\cite{Miyake_etal98}
applied the path--integral Monte Carlo technique to the study of hydrogen and
muonium at the T site in silicon, with the electron-electron interactions
described within the LDA. Despite finding the T site to be a local maximum on
the potential energy surface, they found the muon distribution to be peaked at
that site because of the quantum motion.

In this work we employ all--electron DFT calculations within a double adiabatic
approximation to study muonium and hydrogen at the BC and T sites in silicon
and germanium. The use of all--electron calculations allows an assessment of
the accuracy of the correction procedures which are used to obtain the
hyperfine and superhyperfine parameters in pseudopotential
calculations.~\cite{VdW_Blochl93} The use of a double adiabatic approximation
allows us to obtain both the zero--point energy and wave function of the
impurity. Our inclusion of the zero--point motion is at a level beyond that in
previous first principles calculations since the positions of the host silicon
or germanium atoms are allowed to relax in the presence of the zero--point
motion of the impurity.  At this level of approximation the relaxations of the
host lattice are different for a muon and a proton. Our calculations thus allow
an assessment of the differences in the potentials felt by the two impurities,
thereby testing one of the assumptions underlying the configuration--coordinate
diagram~\cite{config} used to interpret experimental data.

\section{Method}
\label{method}

\subsection{All--electron spin--polarized LSDA--DFT calculations}

\begin{figure}
\caption{The muon/proton at the bond--centred (a) and tetrahedral (b)
sites in silicon.  The dashed circles in (a) show the unrelaxed positions
of the silicon atoms.  These are not shown in (b) because the
relaxation around the T site is negligible.}
\begin{center}
\epsfysize=9cm \epsfbox{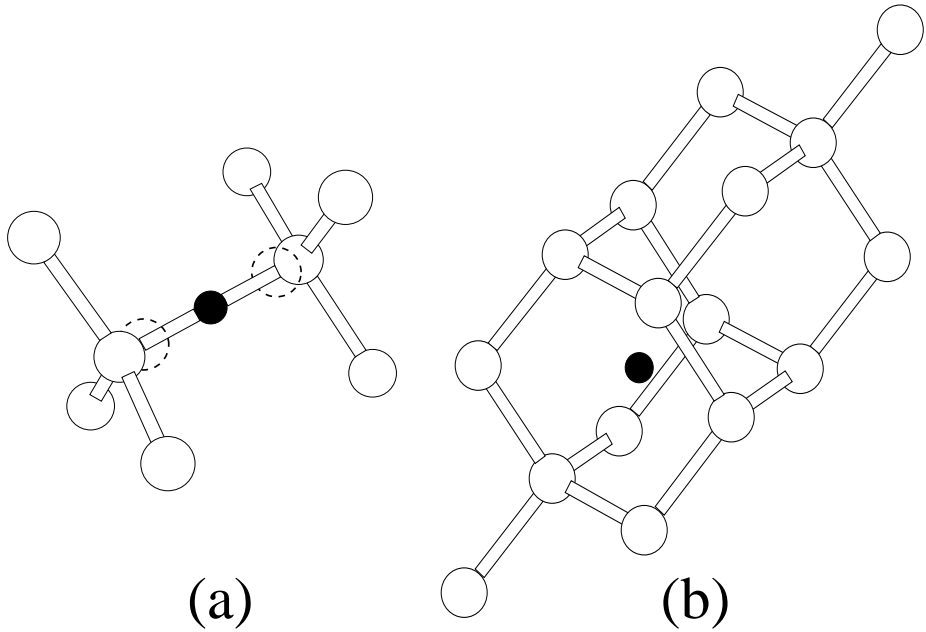}
\end{center}
\label{fig:geometries}
\end{figure}

All the first principles calculations reported here were performed with the
CRYSTAL95 software package~\cite{crystal}. The (zero temperature)
spin--polarized density functional method~\cite{HKtheorem,KohnSham} was used,
together with both local density and gradient corrected approximations to the
exchange--correlation functional (namely the Perdew--Zunger LSDA~\cite{PZ} and
the PW91 form of the GGA~\cite{GGA}). The calculations were performed within a
periodic supercell approach with a single hydrogen impurity in supercells
containing either sixteen or fifty--four silicon or germanium atoms.
Fig.~\ref{fig:geometries} shows the relaxed  atomic environments of a single
muon at both bond--centred and tetrahedral impurity sites in silicon. The
measured lattice constants (5.429 \AA \ for silicon and 5.6579 \AA \ for
germanium) were used in all calculations.

Other approximations made were as follows. The use of local basis functions
requires the real space Coulomb and exchange series to be limited and
approximated as described in references~\onlinecite{crystal,pisani}; the
accuracy with which the various Gaussian integrals are computed is controlled
by classifying basis function pairs according to overlap or penetration
criteria defined by five parameters, which in this study were set to $10^{-7}$,
$10^{-6}$,  $10^{-7}$, $10^{-7}$ and $10^{-14}$.~\cite{crystal} This is
normally sufficient to give a numerical error of less than 0.001~eV/atom in the
relative energies of different structures. The reciprocal space integrations
necessary to reconstruct the density matrix in real space at each
self-consistent cycle were approximated by summing over a set of k--points
belonging to a mesh of Monkhorst--Pack~\cite{Monk_Pack} type which was centred
on the origin in reciprocal space. The convergence of both the total energy and
the isotropic hyperfine parameter of the muon with respect to the reciprocal
space sampling density was investigated.  A $4\!\times\!4\!\times\!4$ k--point
mesh was found to be sufficient for the 16--atom supercell. With this mesh, the
total energy and isotropic hyperfine parameter are within 0.0025~eV/atom and 3
MHz of their fully converged values, respectively.  A $3\!\times\!3\!\times\!3$
k--point mesh was used for the 54--atom supercell, which also gives excellent
convergence. The convergence of various quantities with respect to the
supercell size and basis set is discussed in Section~\ref{results}.

A hydrogen impurity introduces a defect state into the band gap of the host
crystal. Finite supercell sizes give rise to interactions between the defects
in different cells and thus to a small but potentially significant dispersion
in the defect band. This dispersion could lead, for example, to overlap of the
majority spin defect band with the minority spin defect band and/or the silicon
valence/conduction bands. In either case an unphysical conducting state is
formed. This problem is not entirely eliminated even with the use of the
larger 54--atom supercell. However, judicious use of the level--shifting
convergence technique~\cite{crystal,levshift} allows a small decoupling of
unoccupied and occupied states which prevents the system entering a conducting
state. Population analysis of the final self--consistent wave function revealed
that each supercell contained a single extra majority spin electron as expected
on physical grounds. The calculations thus correctly model this aspect of the
behaviour of a single impurity in a large crystal, which is necessary in order
to obtain physical hyperfine and superhyperfine parameters.

\subsection{Gaussian Basis Sets}

The Bloch functions required to expand the Kohn-Sham orbitals in the
solid--state band structure problem are built from periodic arrays of 
atom--centred Gaussian functions. One motivation for the use of such a basis
set is that all electrons in the system may be treated explicitly, allowing the
spin density at and around the nucleus (and hence the hyperfine parameters) to
be calculated directly without resorting to correction procedures of the sort
required in pseudopotential calculations.~\cite{VdW_Blochl93}

The basis set used for the majority of the silicon calculations was of the type
{\it s}(8){\it sp}(8){\it sp}(3){\it sp}(1) where the numbers in brackets refer
to the number of contracted primitive Gaussians making up each shell. For
convergence checking we also used a higher quality silicon set with an
additional {\it d} polarization function of the type {\it s}(8){\it sp}(8){\it
sp}(1){\it sp}(1){\it sp}(1){\it sp}(1){\it d}(1). The basis set used for the
germanium calculations was {\it s}(9){\it sp}(7){\it sp}(6){\it sp}(3){\it
sp}(1){\it d}(6){\it d}(1).

To describe the hydrogen atom, an uncontracted basis of eleven {\it s}
functions and a single {\it p} function was used. Such a large set (mainly
consisting of functions with relatively high exponents) was found to be
necessary to compute accurate hyperfine parameters. A spin--unrestricted
Hartree--Fock calculation of the total energy of the free atom with this basis
gave $-0.49988$~Ha which is close to the exact result of $-0.5$~Ha. The
isotropic hyperfine parameter was 1421.9~MHz compared with that obtained from
the exact wave function of 1422.8~MHz. The corresponding values obtained from
an LSDA--DFT calculation with this basis set were $-0.47833$ Ha (which is very
close to the value of $-0.47885$~Ha obtained from an atomic code using
integration on a very fine grid) and 1356.6~MHz.

Optimal Gaussian basis sets for use in close packed solids are significantly
different from those appropriate to the atomic and molecular cases. In
particular, careful optimization is required to avoid the problems of linear
dependence and basis set superposition error due to the overlap of diffuse
functions. In this study we used the following procedure. All basis set
parameters were first optimized in the free atom. The exponents and contraction
coefficients of the valence functions in silicon and germanium were then
reoptimized in the pure bulk material. Finally, a hydrogen atom was inserted at
a bond--centred site, the positions of the nearest--neighbour silicon/germanium
atoms relaxed, and the parameters of the valence functions of each atom again
reoptimized.
To test the transferability of the optimized basis sets the hydrogen was
displaced from the BC site along the bond by 0.27 \AA, and the basis function
parameters were reoptimized for the new geometry. The energy as a function of
displacement along the bond was calculated for each of these two basis sets.
The variation in energy was essentially the same.

The final exponents and contraction coefficients of all the basis sets employed
in this study are available elsewhere.~\cite{basis_set_library}

It is important to investigate the possibility of basis set superposition error
(BSSE) in defect energetics calculations for a system described by a localized
basis. The basis sets for the host lattice atoms are necessarily incomplete.
Insertion of an impurity atom allows additional variational freedom in the
description of the atoms adjacent to the defect site, particularly when the
impurity is described by a relatively diffuse basis set. This can distort the
relative stabilities of defects at impurity sites of differing local
coordination number and geometry. In the present case, the hydrogen impurity is
considerably closer to its neighbours at the BC site than at the T site, and
thus one might expect the BC site to be artificially stabilized with respect to
the T.

This expectation is confirmed by an estimate of the BSSE using a counterpoise
correction.~\cite{counterpoise} For the 16--atom supercell, addition of
``ghost'' hydrogen basis functions into the relaxed silicon lattice lowered the
energy per cell by 0.199~eV (BC site) and 0.068~eV (T site) with the smaller
silicon basis, and by 0.0533~eV (BC site) and 0.0243~eV (T site) with the
larger silicon basis. In germanium, the energy is lowered by 0.251 eV (BC site)
and 0.0534~eV (T site) by the same procedure. Inclusion of a lattice of
``ghost'' silicon/germanium functions around a hydrogen atom lowered the energy
by less than 0.0005~eV. These numbers may be taken to give a rough indication
of basis set incompleteness in each case. It can thus be concluded that in
silicon the BSSE lowers the energy of the BC site over that of the T site by
around 0.13~eV with the smaller basis, but by only 0.03~eV with the larger set.
The corresponding correction for the germanium case is 0.20~eV.

\subsection{Calculation of zero--point motion}
\label{zero}

For the calculation of the zero--point motion of the muon/proton a double
adiabatic approximation was used. This means that the motions of the electrons
and of the muon are considered to be decoupled from the motion of the atomic
nuclei, and furthermore that the electronic motion is decoupled from that of
the muon. The approximation is justified by the fact that a muon is roughly
207 times more massive than an electron and around 243 times less massive than
a silicon nucleus. For a proton the equivalent factors are respectively 1836
and 28; the decoupling of the proton and silicon motion is thus somewhat less
justified.  The mass differences are of course more favourable for the heavier
germanium nucleus.  

The positions of the silicon/germanium nuclei are denoted by ${\bf
r}_{n}$, the muon or proton positions by ${\bf
r}_{\mu}$, and the electron positions by ${\bf r}_{e}$.  The double
adiabatic approximation is used to decouple the motions of the
particles by approximating the wave function as a product of nuclear,
muon/proton and electronic parts,
\begin{equation}
\Psi({\bf r}_{e},{\bf r}_{\mu},{\bf r}_{n})=\psi^{n}({\bf r}_{n})
X^{\mu}({\bf r}_{\mu};{\bf r}_{n}) \phi^{e}({\bf
r}_{e};{\bf r}_{\mu},{\bf r}_{n}) \ ,
\end{equation}
where the variables to the right of the semi--colons appear as
parameters and those to the left are dynamical variables.  Within the
double adiabatic approximation the three wave functions each satisfy
separate Schr\"{o}dinger equations:
\begin{equation}
\hat{H}_{e}({\bf r}_{e};{\bf r}_{\mu},{\bf r}_{n})\, \phi^{e}({\bf r}_{e};{\bf r}_{\mu},{\bf r}_{n})= E^{e}({\bf r}_{\mu},{\bf r}_{n})\,\phi^{e}({\bf r}_{e};{\bf r}_{\mu},{\bf r}_{n}) \ ,
\label{eq:e_SE}
\end{equation}
\begin{equation}
\hat{H}_{\mu}({\bf r}_{\mu};{\bf r}_{n})
\,X^{\mu}_{\alpha}({\bf r}_{\mu};{\bf r}_{n})=
E^{\mu}_{\alpha}({\bf r}_{n})\,X^{\mu}_{\alpha}({\bf r}_{\mu};{\bf r}_{n}) \
,
\label{eq:mu_SE}
\end{equation}
\begin{equation}
\hat{H}_{n}\psi^{n}_{\alpha}({\bf r}_{n})= E^{n}_{\alpha} \psi^{n}_{\alpha}({\bf
r}_{n}) \ .
\label{eq:I_SE}
\end{equation}
The subscript $\alpha$ labels the different eigenstates of the muon. Although
only the ground state of the nuclear wave function is considered here, it is
also labelled by $\alpha$ since it depends on the chosen muon eigenstate. The
different electronic eigenstates are not labelled, since it is only the ground
state of the electronic wave function as a function of the muon and nuclear
positions that is of interest in the current work. The three Hamiltonians are:
\begin{eqnarray}
\hat{H}_{e}({\bf r}_{e};{\bf r}_{\mu},{\bf r}_{n})&= &\hat{T}_{e}({\bf r}_{e})+V_{ee}({\bf r}_{e}) \\ \nonumber  
&+&V_{en}({\bf r}_{e},{\bf r}_{n})+V_{e\mu}({\bf
r}_{e},{\bf r}_{\mu}) \ ,
\label{eq:H_e}
\end{eqnarray}
\begin{eqnarray}
\hat{H}_{\mu}({\bf r}_{\mu};{\bf r}_{n})&=&\hat{T}_{\mu}({\bf r}_{\mu})+V_{\mu\mu}({\bf r}_{\mu})+ V_{\mu n}({\bf r}_{\mu},{\bf r}_{n}) \\ \nonumber
&+&E^{e}({\bf r}_{\mu},{\bf r}_{n}) \ ,
\label{eq:H_mu}
\end{eqnarray}
\begin{equation}
\hat{H}_{n}({\bf r}_{n})=
V_{nn}({\bf r}_{n}) +
E^{\mu}_{\alpha}({\bf r}_{n}) \ .
\label{eq:H_I}
\end{equation}
where $\hat{T}$ is the kinetic operator and $V_{ab}$ is the Ewald interaction
between particles of types {\it a} and {\it b}.  The term $V_{\mu\mu}$ is a
constant describing the interactions between the impurity atoms in the
different supercells.  The electronic energy, $E^{e}$, appears as an effective
potential in the muonic Hamiltonian, Eq.~\ref{eq:H_mu}. Hence, when
Eq.~\ref{eq:mu_SE} is solved, the resulting energy includes the electronic
contribution.  This energy then appears in the nuclear Hamiltonian as an
effective potential.  Thus the total energy of the system is given by the
eigenvalue in Eq.~\ref{eq:I_SE}.

In order to find a good starting point for the BC calculations we performed
LSDA calculations with the muon/proton fixed at the bond centre. The positions
of the nearest and next--nearest neighbour (NN and NNN) silicon/germanium atoms
were then relaxed.  For the T site the muon/proton was held fixed at the T site
while the four NN silicon/germanium atoms were relaxed in the radial direction.

The next step is to calculate the potential experienced by the muon/proton in
the crystal by solving the electronic Schr\"{o}dinger equation (~\ref{eq:e_SE})
within the LSDA, as a function of the parameters ${\bf r}_{\mu}$ and ${\bf
r}_{n}$.  For the BC site calculations, the parameters ${\bf r}_{n}$ were
varied by considering four different additional relaxations of the NN silicon
atoms along the [1 1 1] direction.  For each of the positions of the NN silicon
atoms, the NNN atoms were relaxed.  We then performed a further twelve LSDA
calculations as a function of the position of the muon/proton for each of these
nuclear configurations in order to map out the required potential energy
surfaces.  A similar procedure was used for germanium. For the T site the
static relaxation of the NN atoms is very small and we assumed that the
zero--point motion of the muon/proton would not give any additional relaxation.

In order to solve Eq.~\ref{eq:mu_SE} for the muon/proton wave function we used
a fitted polynomial for the energy $E^{e}({\bf r}_{\mu},{\bf
r}_{n})$.  For the BC site we use a cylindrical coordinate
system with the origin at the BC site and the $z$-- and $\rho$--coordinates
directed along the bond and in the plane perpendicular to the bond,
respectively.  We have neglected the $\theta$ dependence of the potential. 
This assumption was checked by displacing the muon by 0.53 \AA \ from the BC
site along the [-1 1 0] direction and then rotating it about the [1 1 1]
axis.  The maximum variation seen in the energy of the 16--atom supercell
during the rotation was just 0.002~eV.  The Taylor expansion of the
cylindrically symmetric potential, neglecting sixth--order terms and higher, is

\begin{equation} V_{\rm BC}(\rho,z)=V_{\rm BC}(0,0) + \beta \rho^{2} + \gamma
z^{2} + \delta \rho^2z^2 + \zeta \rho^{4} + \eta z^{4} \ . \label{eq:BCexpn}
\end{equation}

 As a further simplification in order to avoid costly, low--symmetry
calculations, the $\delta \rho^2z^2$ term was neglected.  The resulting
Schr\"odinger equation is separable.  In order to check the assumption of
$\rho$--$z$ separability, a few calculations were performed with the muon at
points where both the $\rho$ and $z$ coordinates were non--zero.  These
energies were then compared with those predicted by the fitted potential
neglecting the term in $\rho^2 z^2$.  The errors due to neglecting the $\rho^2
z^2$ term increased only slowly away from the BC site, and the corresponding
error in the ground state energy is small because the wave function of the
muon/proton is localized around the BC site.

Our approximation formula was then obtained by a least--squares fit of the
twelve energies calculated at different values of ${\bf r}_{\mu}$ (for fixed
${\bf r}_{n}$) to the resulting polynomial.  The fitted values of the
parameters in Eq.~\ref{eq:BCexpn} used in the calculations of the
zero--point energies for silicon and germanium are given in
Table~\ref{tble:BC_param}.

The polynomial expansion for the energy surface around the T site is the Taylor
expansion which is invariant under all of the 24 rotations forming the group
$T_d$.  Cartesian coordinates centred at the T site were used, and in order to
obtain a good fit to the energy surface it was found necessary to include terms
up to sixth order:
\begin{eqnarray}
V_{\rm T}(x,y,z)& = & V_{\rm T}(0,0,0) + a_{2}(x^2 + y^2 + z^2) +
a_{3}\, xyz \nonumber \\
&+& a_{4}(x^4 + y^4 + z^4) + b_{4}(x^{2}y^{2} + x^{2}z^{2} + y^{2}z^{2}) \nonumber \\
&+& a_{5}\,xyz(x^2 + y^2 +
z^2) + a_{6}(x^{6}+y^{6}+z^{6}) \nonumber \\ \hbox{} & + &
b_{6}(x^{2}y^{4} + x^{2}z^{4} + x^{4}y^{2} + x^{4}z^{2} + y^{2}z^{4} \nonumber \\ &&+y^{4}z^{2}) + c_{6}\,x^{2}y^{2}z^{2} \ .
\label{eq:Texpn}
\end{eqnarray}
The fitted values of the parameters in this equation are given in
Table~\ref{tble:T_param}.

The solution of the muon/proton Schr\"{o}dinger equation (~\ref{eq:mu_SE}) was performed by diagonalizing within a basis of
harmonic oscillator eigenfunctions centred on either the BC or T site.
We constructed muon and proton basis sets consisting of the solutions
of the harmonic part of the calculated potentials:
\begin{equation}
V_{0_{\rm BC}}= V_{\rm BC}(0,0,0) + \beta(x^{2} + y^{2}) + \gamma z^{2} \ ,
\end{equation}
and
\begin{equation}
V_{0_{\rm T}} = V_{\rm T}(0,0,0) + a_{2}(x^2 + y^2 + z^2) \ .
\end{equation}
For a harmonic potential, the Schr\"{o}dinger equation may be solved
by separation of variables:
\begin{eqnarray}
X({{\mathbf r}_{\mu}}) & = & {\mathcal X}(x){\mathcal Y}(y){\mathcal Z}(z) \
. \nonumber
\end{eqnarray}
This gives three separate equations for $\mathcal{X}$, $\mathcal{Y}$,
and $\mathcal{Z}$, whose solutions are of the form:
\begin{equation}
{\mathcal X}_{l}(x^{\prime})=A_{l}H_{l}(x^{\prime})e^{-\frac{1}{2}
x^{\prime 2}},
\end{equation}
where $A_{l}=\frac{1}{\sqrt{2^{l}\pi^{\frac{1}{2}}l!}}$ is the
normalisation factor, $x^{\prime}=(2MC_{x})^{\frac{1}{4}}x$ is a
rescaled variable which allows the eigenfunctions to be written in
terms of the standard Hermite polynomials, $H_{l}(x^{\prime})$, and
$C_{x}$ is the appropriate harmonic coefficient.  The associated
energy eigenvalues are
\begin{equation}
E_{\lambda_{l}}=\left(l+\frac{1}{2}\right) \left( \frac{2 C_{\lambda}}{M}
\right)^{\frac{1}{2}} \ ,
\end{equation}
where $\lambda$ runs over the three directions, $x$, $y$, and $z$, and
$C_{\lambda}$ is the harmonic coefficient corresponding to that
direction.  The value of the mass, $M$, of the particle depends on
whether we are solving for the proton or muon wave function.

Having constructed our basis functions we now consider the full
potentials which are written as the sum of harmonic and anharmonic
terms:
\begin{eqnarray}
V_{\rm BC} & = & V_{0_{\rm BC}} + \Delta V_{\rm BC} \\
V_{\rm T}  & = & V_{0_{\rm T}} +  \Delta V_{\rm T} \ .
\end{eqnarray}
The Hamiltonian matrix elements were calculated in the basis of the harmonic
solutions and the resulting matrix equations were diagonalized.  A basis set
constructed from all Hermite polynomials up to and including eighth order and
containing a total of 729 basis functions was found to be sufficient to obtain
converged values for at least the lowest six eigenvalues,
$E^{\mu}_{\alpha}({\bf r}_{n})$, of the system with the muon/proton at the BC
site.  At the T site it was found necessary to increase the size of the basis
set to include all Hermite polynomials up to twelfth order, which gave 2197
functions.

The solution of Eq.~\ref{eq:H_I} is trivial as the operator is
multiplicative and the eigenfunctions are delta functions.  The total
energy, $E^{n}_{\alpha}$, is therefore the sum of
$E^{\mu}_{\alpha}({\bf r}_{n})$ and the Ewald
energy of the lattice of host atoms, $V_{nn}({\bf
r}_{n})$.

\subsection{Hyperfine and superhyperfine parameters and motion averaging}
\label{hyperfine_mthd}

The components of the hyperfine tensor, ${\bf A}$, define the spin Hamiltonian
for the hyperfine interaction between the spins of an electron and a nucleus:

\begin{equation}
\hat{\mathcal H}_{s}={\bf S}_{e} \cdot {\bf A} \cdot {\bf
S}_{n} \ .
\end{equation}
The hyperfine tensor is normally split into isotropic and anisotropic
parts,
\begin{equation}
{\bf A} = A_s{\bf I} + {\bf A}_{p} \ ,
\end{equation}
\noindent where ${\bf I}$ is the ($3\times3$) unit matrix.  The isotropic
hyperfine parameter (or {\it superhyperfine} parameter when it is
calculated at one of the nearest neighbours of the impurity) is given by
\begin{eqnarray}
A_s & = & \frac{2\mu_{0}}{3}g_{e}\mu^{e}g_{n}
\mu^{n}\rho_{\sigma} ({\mathbf
r}_{n}) \nonumber \\ & = &
104.982\gamma^{n} \rho_{\sigma}({\mathbf
r}_{n}) \ {\mathrm [MHz]} \ .
\end{eqnarray}
where $\mu_{0}$ is the permeability of free space, $\mu^{e}$ is the
Bohr magneton, $\mu^{n}$ is the nuclear magneton
and $g_{e}$ and $g_{n}$ are the electron and
nuclear {\it g} factors.~\cite{footnote} The position of the nucleus
is denoted by ${\mathbf r}_{n}$ and $\rho_{\sigma}
=\rho_{\uparrow} - \rho_{\downarrow}$ [bohr$^{-3}$].~\cite{footnote_2}

The anisotropic part of the hyperfine tensor is given by
\begin{eqnarray}
{\bf A}_{p} & = & \frac{\mu_{0}}{4\pi}g_{e}\mu^{e}g_{n}
\mu^{n} \int {\bf T}({\bf r})
\rho_{\sigma}({\bf r} + {\bf r}_{n})
d{\bf r} \\ \nonumber 
 & = & 12.531\gamma^{n} \int {\bf T}({\bf r})
\rho_{\sigma}({\bf r} + {\bf r}_{n})
d{\bf r}  \ {\mathrm [MHz]} \nonumber \,
\end{eqnarray}
where ${\bf T}({\bf r})$ is a traceless tensor,
\begin{equation}
\label{eq:T}
{\bf T}({\bf r}) = \frac{1}{r^5} \left( \begin{array}{ccc}
3x^{2} - r^{2} & 3xy            & 3xz \\
3xy            & 3y^{2} - r^{2} & 3yz \\
3xz            & 3yz            & 3z^{2} - r^{2}
\end{array} \right) \ ,
\end{equation}
\noindent and the origin of coordinates is on the nucleus at ${\bf
r}_{n}$.  For a particle located precisely at the
BC site the hyperfine interaction has axial symmetry with respect to
the [1 1 1] axis and thus ${\bf A}_{p}$ has the form
\begin{equation}
{\bf A}_{p} = A_{p} \left( 
\begin{array}{c@{\hspace{5mm}}c@{\hspace{5mm}}c}
0 & 1 & 1 \\
1 & 0 & 1 \\
1 & 1 & 0
\end{array} \right) \ ,
\end{equation}
with $A_{p}$ being the anisotropic hyperfine parameter at this site.
For a particle located precisely at the T site, all elements of ${\bf
A}_{p}$ are zero and hence the hyperfine tensor is purely isotropic.

In reality the muon/proton will explore the environment around these sites by
virtue of its zero--point motion and thermal effects.  In order to account for
the zero--point motion the hyperfine interaction tensor must be averaged over
the squared modulus of the muon/proton wave function:
\begin{equation}
\label{eqn:mtn_avrge}
\langle {\bf A} \rangle_{\mu} = \int |X({{\mathbf r}_{\mu}};{\mathbf
r}_{n})|^2 {\bf A}({{\mathbf r}_{\mu}}) d{{\mathbf
r}_{\mu}} \ .
\end{equation}
To evaluate the integral for each component of ${\bf A}$ we fit
$A_s(x_{\mu},y_{\mu},z_{\mu})$ and each of the six distinct elements of the
symmetric tensor ${\bf A}_p(x_{\mu},y_{\mu},z_{\mu})$ to polynomial expressions
of the correct symmetry.  Since the muon/proton wave function is expanded in
terms of Hermite polynomials, analytic expressions for the elements of
$\langle{\bf A}\rangle_{\mu}$ may be obtained.

For the isotropic hyperfine parameters the polynomial expression for
$A_s(x_{\mu},y_{\mu},z_{\mu})$ has the same symmetry as the relevant potential
energy surface.  These parameters were expanded in sixth--order polynomials. 
The polynomial describing the isotropic superhyperfine parameter at the BC site
contains terms which are odd in $z_{\mu}$. (Superhyperfine parameters were not
calculated for the T site.)  Each of the elements of ${\bf A}_{p}^{\rm T}$ and
${\bf A}_{p}^{\rm BC}$ were fitted to second--order polynomials of the correct
symmetry.

We now consider the symmetry of the hyperfine tensor including the effects of
zero--point motion.  Within the double adiabatic approximation the muon motion
is described by the wave function $X({\bf r}_{\mu};{\bf r}_{n})$.  The
motion--average of the $\alpha\beta$--component ($\alpha,\beta = x,y,z$) of the
anisotropic hyperfine tensor is given by:

\begin{equation}
\label{eq1}
\langle A_{\alpha\beta}\rangle_{\mu} = C \int \int |X({\bf r}_{\mu};
{\bf r}_{n})|^2 \rho_{\sigma}({\bf r}+ {\bf
r}_{\mu}) \, T_{\alpha\beta}({\bf r}) \, d{\bf r} \, d{\bf r}_{\mu} \;,
\end{equation}

\noindent where $C$ is a constant, $\rho_{\sigma}$ is the electron
spin density, and $T_{\alpha\beta} $ denotes the components of the
tensor ${\bf T}$ defined in Eq.~\ref{eq:T}.  

The muon/proton may be said to be trapped in a potential well if its wave
function is negligibly small outside of an equi--potential--energy surface
enclosing the region.  The muon wave function, $X$, is the non--degenerate
ground state of the potential well and therefore has the full point--group
symmetry of the well, {\it i.e.},

\begin{equation}
\label{eq2}
P(Q_i)X({\bf r}_{\mu};{\bf r}_{n}) = X({\bf
R}^{-1}_i{\bf r}_{\mu};{\bf r}_{n}) = X({\bf
r}_{\mu};{\bf r}_{n})\;,
\end{equation}

\noindent where $P$ is a scalar transformation operator, $Q_i$ is an
operation of the point group of the well, and ${\bf R}_i$ is the
corresponding transformation matrix.  The electron spin density,
$\rho_{\sigma}({\bf r}+{\bf r}_{\mu})$, satisfies

\begin{equation}
\label{eq3}
P(Q_i) \rho_{\sigma}({\bf r}+{\bf r}_{\mu}) = \rho_{\sigma}({\bf
R}^{-1}_i({\bf r}+{\bf r}_{\mu})) = \rho_{\sigma}({\bf r}+{\bf
r}_{\mu}) \;.
\end{equation}

\noindent $\langle A_{\alpha\beta}\rangle_{\mu}$ is unchanged by a
scalar transformation of the integrand, {\it i.e.},

\begin{eqnarray}
\label{eq4}
\langle A_{\alpha\beta}\rangle_{\mu} = C \int \int P(Q_i) \left[|X({\bf
r}_{\mu}; {\bf r}_{n})|^2 \times \right. &&\nonumber \\
 \left. \rho_{\sigma}({\bf
r}+{\bf r}_{\mu}) T_{\alpha\beta}({\bf r}) \right] \, d{\bf r} \,
d{\bf r}_{\mu} \;. &&
\end{eqnarray}

\noindent $\langle A_{\alpha\beta}\rangle_{\mu}$ is again unaltered if
we sum over the $i$ operations and divide by their number, $N$,

\begin{eqnarray}
\label{eq5}
\langle A_{\alpha\beta}\rangle_{\mu} = \frac{C}{N}\sum_i \int \int
P(Q_i)\left[|X({\bf r}_{\mu}; {\bf r}_{n})|^2
 \times \right. &&\nonumber \\
 \left. \rho_{\sigma}({\bf r}+{\bf r}_{\mu})T_{\alpha\beta}({\bf r})
\right] \, d{\bf r} \, d{\bf r}_{\mu} \;. &&
\end{eqnarray}

\noindent Using Eqs.~\ref{eq2} and~\ref{eq3} we have 

\begin{eqnarray}
\label{eq6}
\langle A_{\alpha\beta}\rangle_{\mu} = \frac{C}{N} \int \int |X({\bf
r}_{\mu}; {\bf r}_{n})|^2 \rho_{\sigma}({\bf
r}+{\bf r}_{\mu}) \times && \nonumber \\
\left[ \sum_i P(Q_i) \, T_{\alpha\beta}({\bf r})
\right] \, d{\bf r} \, d{\bf r}_{\mu}\;. &&
\end{eqnarray}

The symmetry properties of $\langle A_{\alpha\beta}\rangle_{\mu}$ are
easily obtained from Eq.~\ref{eq6}.  For example, $\langle
A_{xy}\rangle_{\mu}$ will be equal to $\langle A_{yz}\rangle_{\mu}$ if
$\sum_i P(Q_i)\,xy = \sum_i P(Q_i)\,yz$.  Taking the specific
case of the T site, it is easily shown that $\sum_i P(Q_i)\,xy =
\sum_i P(Q_i)\,yz = 0$, where the sum is over the 24 operations of the
tetrahedral point group.  Similar arguments show that all the elements
of $\langle A_{\alpha\beta}\rangle_{\mu}$ are zero for the T site.
Similarly, for the BC site we find that all off--diagonal elements of
$\langle A_{\alpha\beta}\rangle_{\mu}$ are equal, and the diagonal
elements are zero.  

If the zero--point motion of the muon is neglected then $|X({\bf
r}_{\mu}; {\bf r}_{n})|^2 = \delta({\bf
r}_{\mu}-{\bf r}_0)$, where ${\bf r}_0$ is the position of the muon.
It follows that if the muon is placed at an invariant point of the
symmetry group of the well, then including the zero--point motion does
not change the symmetry of the anisotropic hyperfine tensor.  This
result explains why the zero--point motion does not affect the
symmetry of the anisotropic hyperfine tensor for either the T or BC
sites considered here.

The presence of the muon could lead to a symmetry lowering distortion
of the host lattice, in which case the appropriate point group is the
lower symmetry one.  We have not considered the possibility of
symmetry lowering distortions in our calculations because of the
computational cost of evaluating the energy $E^{e}({\bf r}_{\mu},{\bf
r}_{n})$ of Eq.~\ref{eq:e_SE} for the required
atomic configurations.  However, we believe such distortions to be
unlikely for the cases considered here.

\section{Results}
\label{results}

\subsection{Static relaxations}
\label{static}

The static relaxations (neglecting zero--point motion) are, of course,
identical for the muon and proton.  Calculations with the muon/proton fixed at
the BC site of the 16--atom silicon cell showed that the two nearest neighbours
of the muon/proton relax outwards from the muon/proton by 0.40 \AA \ along the
[1 1 1] axis with the NNNs relaxing by 0.01 \AA \ in the same direction.  The
corresponding relaxations for the 54--atom supercell were 0.39 \AA\ and 0.02
\AA. These values are close to the plane--wave pseudopotential results of
Luchsinger {\it et al.}~\cite{Luch97} who obtained relaxations of 0.45 \AA \
for the NN silicon atoms and 0.07 \AA \ for the NNNs.  For the 16--atom
germanium cell the corresponding relaxations were 0.44 \AA \ for the NNs and
0.02 \AA \ for the NNNs.  Our NN relaxation is in good agreement with the value
of 0.42 \AA \ calculated by Vogel {\it et al.}~\cite{Vogel89}

At the T site, the NN atoms in the host lattice were allowed to relax in the
radial direction.  The relaxations in silicon and germanium were approximately
equal and very small; just 0.02 \AA \ {\it towards} the muon/proton in the
16--atom supercell and 0.03 \AA \ (in the same direction) in the 54--atom
supercell.  Again, this is in agreement with the ``negligible'' relaxation for
the T site in silicon found by Luchsinger {\it et al.}

\subsection{Relaxations including zero--point motion}
\label{dynamic}

The influence of the zero--point motion of the muon/proton on the relaxation of
the silicon/germanium host lattice was studied by calculating the total energy,
$E^{n}_{\alpha}$, of Eq.~\ref{eq:I_SE}, for different
relaxations of the NN host atoms, as described in Section~\ref{zero}.

\begin{figure}
\caption{The square of the proton (thin solid line) and muon (dashed
line) wave functions at the BC site in silicon. The symbols denote the
potential well for the 16-- and 54--atom supercells, and the thick
solid line is the fit of the 16--atom data to
Eq.~\protect\ref{eq:BCexpn} with the parameters of
Table~\protect\ref{tble:BC_param}.}
\begin{center}
\epsfysize=8cm \epsfbox{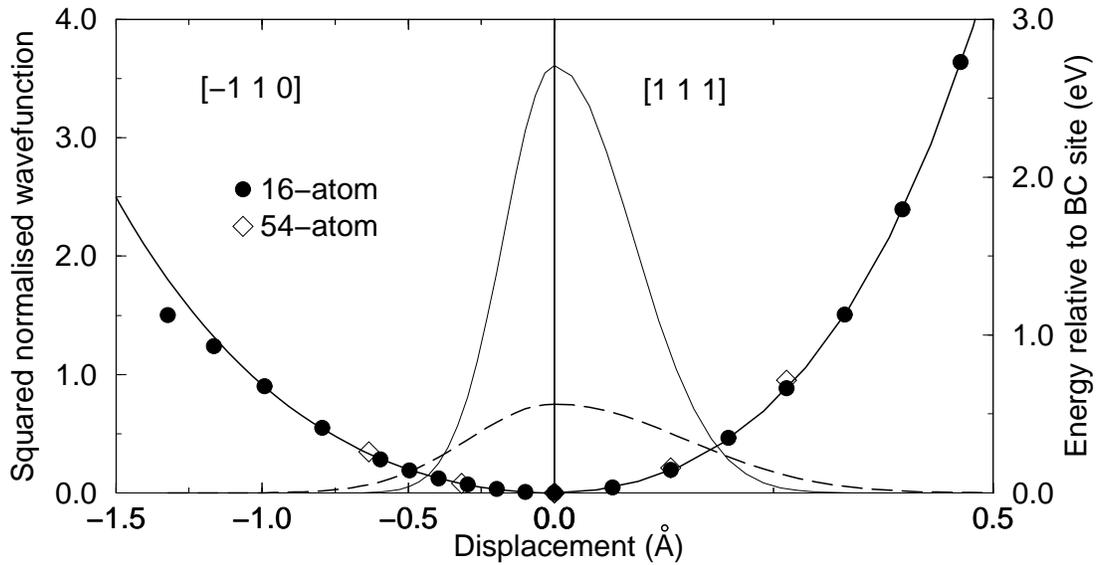}
\end{center}
\label{fig:finsize_pe}
\end{figure}

For the BC site four different relaxations of the NNs were considered, and for
each of these the six NNNs were also relaxed. The NN relaxations are in
addition to the static relaxations given in Section~\ref{static}. The inclusion
of the zero--point energy of the muon was found to give only a small correction
to the static relaxations; the NN silicon atoms relaxed outwards by just an
additional 0.01 \AA \ in the [1 1 1] direction, so that the final separation of
the muon from a NN silicon atom is 1.58 \AA \ in the 16--atom supercell.  The
much smaller zero--point energy of the heavier hydrogen impurity is swamped by
the increase in energy of the crystal as the separation of the NN atoms is
increased and thus there is no additional relaxation.  As a check on the
finite--size errors, the energies of five geometries were recalculated using
the 54--atom supercell.  These energies and the corresponding potential energy
curves calculated within the 16--atom supercell are shown in
Fig.~\ref{fig:finsize_pe}.  The very small differences between the 16--atom and
54--atom results justify the use of the 16--atom supercell in calculations of
the shape of the potential well at the BC site.

\begin{figure}
\caption{The square of the proton (thin solid line) and muon (dashed
line) wave functions at the BC site in germanium. The symbols denote
the potential well for the 16--atom supercell, and the thick solid
line is the fit to Eq.~\protect\ref{eq:BCexpn} with the parameters of
Table~\protect\ref{tble:BC_param}.}
\begin{center}
\epsfysize=8cm \epsfbox{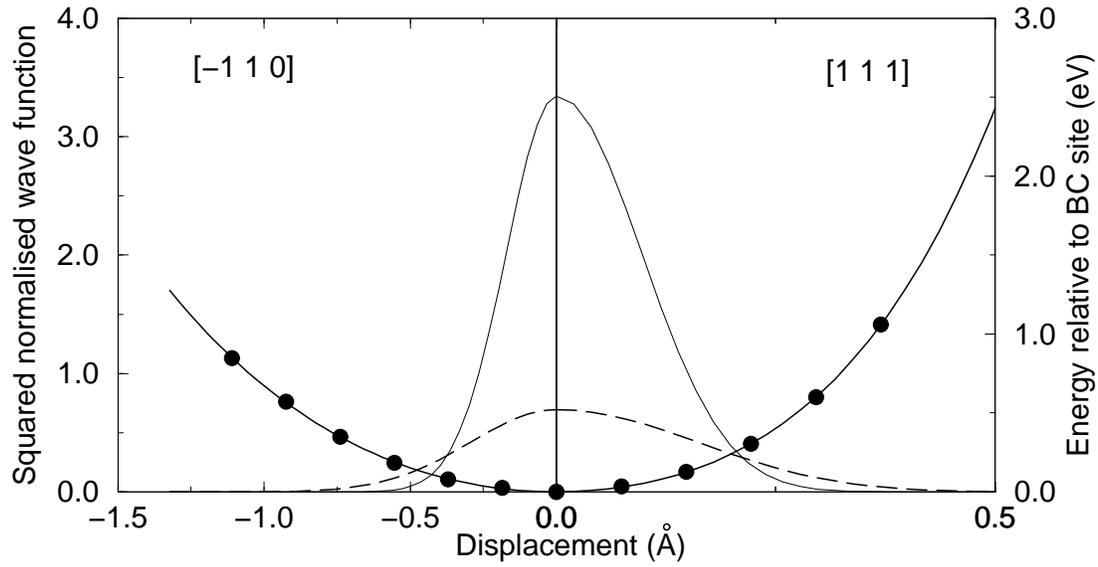}
\end{center}
\label{fig:bc_wfns}
\end{figure}

The story is very similar for the BC site in germanium.  When the zero--point
energy of the muon is included, the relaxation of the NN atoms again increases
by just 0.01 \AA, so that the final separation of the muon from a NN germanium
atom is 1.69 \AA \ in the 16--atom supercell.  Once more, the smaller
zero--point energy of the proton means there is no additional relaxation due to
quantum effects. Fig.~\ref{fig:bc_wfns} shows the potential energy well and
calculated wave functions for the muon and proton at the BC site in germanium.

\subsection{Zero--point energies}
\label{zp_energies}

\begin{figure}
\caption{The calculated potential experienced by the muon/proton in the
$\{110\}$ planes for the BC site in silicon.  The figure shows
both of the NN and two of the NNN silicon atoms of the muon/proton
with the bond lengths drawn to scale.  The contours range from 0.1 to
0.7~eV in increments of 0.1~eV.}
\begin{center}
\epsfysize=9cm \epsfbox{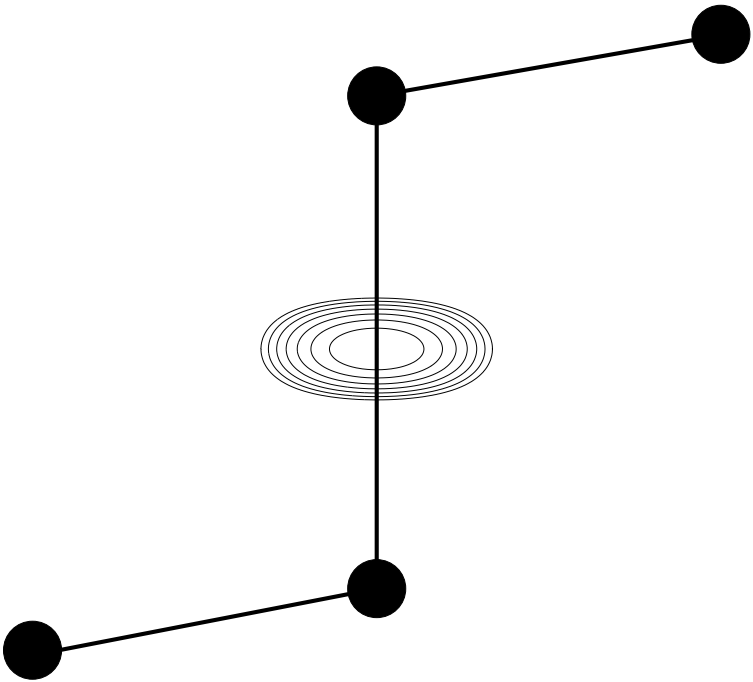}
\end{center}
\label{fig:bc_pot}
\end{figure}

The zero--point energy of the muon at the BC site was calculated to be 0.63~eV
in silicon and 0.56~eV in germanium.  It is perhaps surprising that such large
zero--point energies have so little effect on the relaxations.  As shown in
Fig.~\ref{fig:bc_pot}, the potential well is narrow in the direction along the
bond and wider perpendicular to the bond.  Within the harmonic approximation
one can decompose the zero--point energy into contributions from the well along
and perpendicular to the bond.  For a muon at the BC site in silicon this gives
0.47~eV in the direction along the bond and 0.22~eV perpendicular to the bond. 
(The sum of these differs from the full zero--point energy of 0.63~eV because
the latter does not assume the harmonic approximation.)  The corresponding
energies for germanium are 0.37~eV and 0.22~eV in the directions along and
perpendicular to the bond, respectively.  If we were to consider only the
zero--point energy in the direction along the bond then the outwards relaxation
of the silicon/germanium atoms would be larger; approximately 0.03 \AA \ in
silicon and more than 0.025 \AA \ in germanium.  Although the component of the
zero--point energy along the bond is significantly reduced by further outward
relaxation of the silicon/germanium atoms, the potential well also gets
narrower in the plane perpendicular to the bond, which tends to increase the
zero--point energy.  The narrowing of the potential well in the plane
perpendicular to the bond correlates with the narrowing of the bonding charge
cloud as the bond lengthens.

Our result of 0.63~eV for the zero--point energy of the muon at the BC site in
silicon is close to the value of 0.54~eV obtained by Claxton {\it et
al.}~\cite{Claxton92} from Hartree--Fock calculations on Si$_{26}$H$_{30}$
clusters.  In that calculation the potential well at the BC site was assumed to
be cylindrically symmetric about the bond (as it is in this work) and the
resulting Schr\"{o}dinger equation was solved within the harmonic
approximation.

The larger mass of the proton significantly reduces the quantum effects.  We
calculated the zero--point energy of a proton at the BC site to be 0.20~eV in
silicon and 0.18~eV in germanium.  Our value for silicon is close to that of
0.18~eV obtained by Luchsinger {\it et al.}~\cite{Luch97}, which suggests that
the harmonic approximation to the potential well used in that work is quite
good for the proton ground state.

\begin{figure}
\caption{The square of the proton (thin solid line) and muon (dashed
line) wave functions at the T site in silicon. The symbols denote the
potential well for the 54--atom supercell, and the thick solid line is
the fit to Eq.~\protect\ref{eq:Texpn} with the parameters of
Table~\protect\ref{tble:T_param}. The potential has a maximum at the
hexagonal site situated at 1.18 \AA \ from the T site, but the fit has
been constrained so that it forms a simple potential well.}
\begin{center}
\epsfysize=9cm \epsfbox{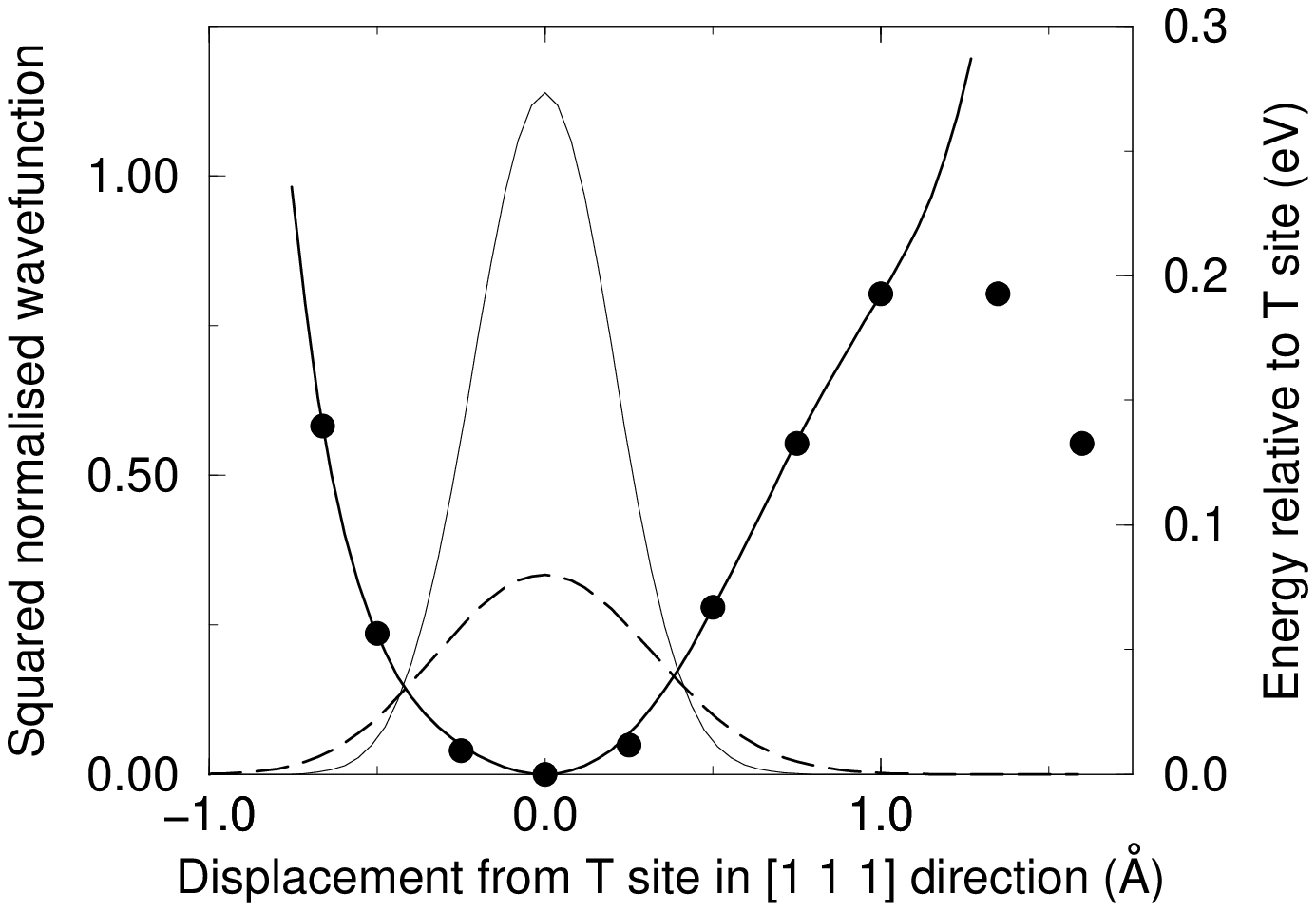}
\end{center}
\label{fig:T_si_wfns}
\end{figure}

In contrast to the results of plane--wave pseudopotential
calculations~\cite{Luch97,VdW_Dent89}, we find the T site corresponds to a
local minimum in the potential energy surface.  This is, however, in agreement
with a more recent plane--wave pseudopotential study.~\cite{Probert} The
calculated energy surface along the [1 1 1] direction is shown in
Fig.~\ref{fig:T_si_wfns}.  It turns out that the muon/proton is not strongly
bound in our potential well, which turns over at the hexagonal site situated at
a distance of 1.18 \AA\ from the T site along the [1 1 1] direction.  To
confine the muon/proton in the well we therefore constrained the fit to prevent
the potential turning over, as shown in Fig.~\ref{fig:T_si_wfns}.  The
zero--point energy of the muon/proton calculated in such a well is then an
upper bound on the true value, but as the wave function decays quite rapidly
away from the T site this bound is accurate.

In contrast to the BC site, investigation of the finite size effects present in
the calculation of this energy surface showed that while the results in the
16-- and 54--atom supercells were qualitatively similar, they differed
significantly in the openness and depth of the potential well.  As a result,
the calculation of the zero--point energy {\it etc.} of the muon/proton at this
site was carried out using the potential well obtained from the 54--atom
supercell.  The expense of the LSDA calculations with this supercell made the
generation of data points off the [1 1 1] axis too costly. Therefore, the three
parameters left undetermined after the one--dimensional fit to the data on the
[1 1 1] axis were assigned the values obtained in the fit to the 16--atom
supercell data.  This is not a critical choice since the one--dimensional fit
has already constrained the shape of the energy surface in eight directions
(due to the symmetry of the T site).  This procedure was also used to generate
a three--dimensional potential--energy function from the fit to data along the
[1 1 1] axis (calculated using the 16--atom supercell) at the T site in
germanium.  The parameters obtained from these fits are given in
Table~\ref{tble:T_param}.

\begin{figure}
\caption{The square of the proton (thin solid line) and muon (dashed
line) wave functions at the T site in germanium. The symbols denote the
potential well for the 16--atom supercell, and the thick solid line is
the fit to Eq.~\protect\ref{eq:Texpn} with the parameters of
Table~\protect\ref{tble:T_param}. The potential has a maximum at the
hexagonal site situated at 1.23 \AA \ from the T site, but the fit has
been constrained so that it forms a simple potential well.}
\begin{center}
\epsfysize=9cm \epsfbox{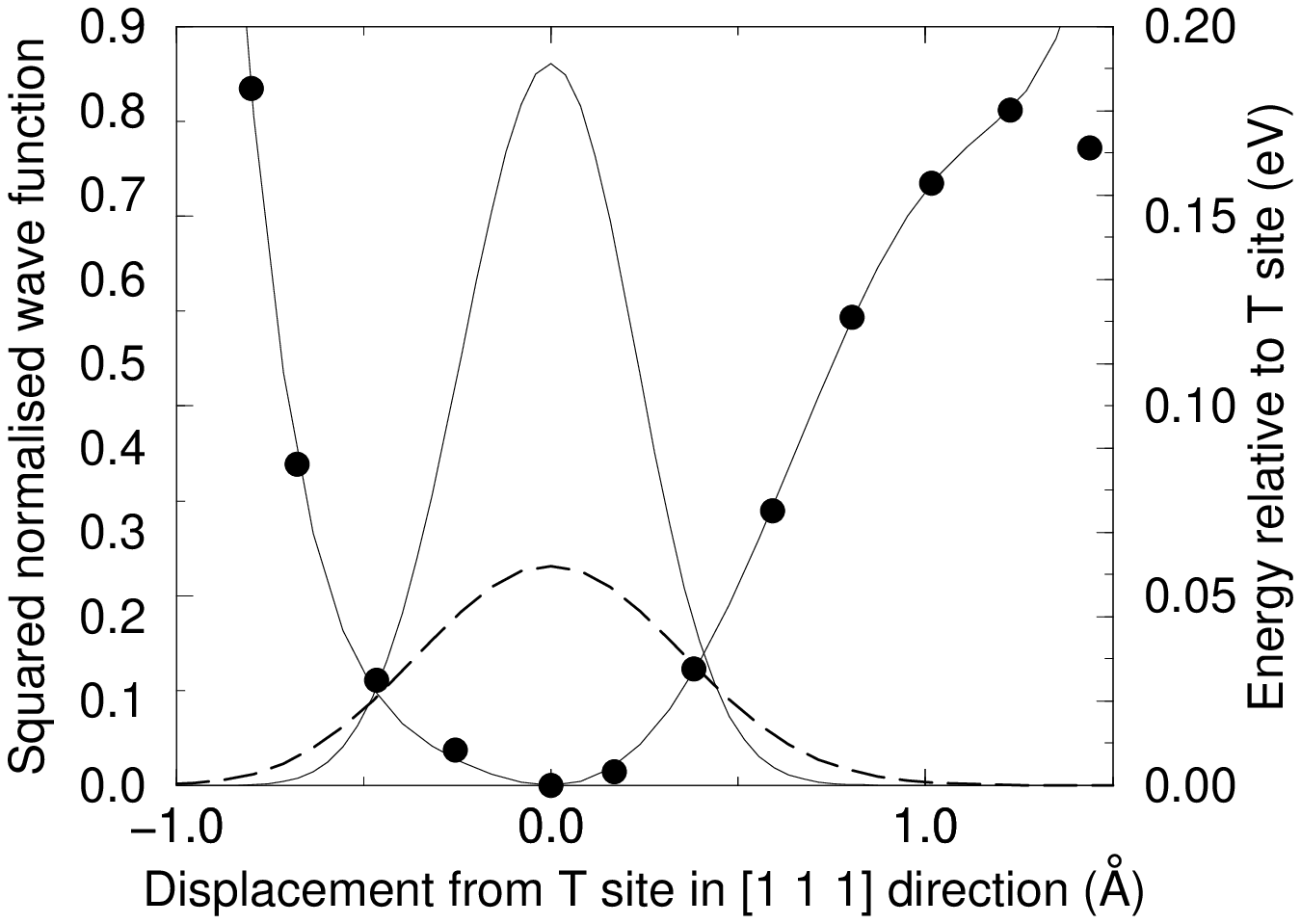}
\end{center}
\label{fig:T_ge_wfns}
\end{figure}

For a muon at the T site the zero--point energy was calculated to be 0.28~eV in
silicon and 0.22~eV in germanium. For a proton the corresponding values are
0.09~eV and 0.06~eV.  The ground state wave functions of the muon/proton along
the [1 1 1] axis in silicon are shown in Fig.~\ref{fig:T_si_wfns} and those in
germanium in Fig.~\ref{fig:T_ge_wfns}.  The results for germanium must be
considered approximate since we have not calculated any data points off the [1
1 1] axis in this case.  In addition, the 16--atom supercell was used for all
of the germanium calculations and therefore it follows from the behaviour found
in silicon that the true potential energy surface will be more open than the
one we have obtained.

\subsection{Excited states of the muon and proton}
\label{excited}

For a muon at the BC site in silicon our zero--point energy of 0.63~eV is
considerably smaller than the well depth of 1.37~eV, indicating the possibility
that excited states of the muon may be bound within the well.  Numerical
calculations show a two--fold degenerate first excited state at an energy of
0.84~eV.  The wave functions of these states are similar to those obtained from
a harmonic approximation, {\it i.e.}, they consist essentially of an excitation
within the plane perpendicular to the bond.  The energy of the first excited
state is also reasonably well described within the harmonic approximation which
predicts the excited state to be 0.22~eV above the ground state.  It is also
possible that some of the higher energy states are bound within the well.  In
germanium, the potential well at the BC site is 1.51~eV deep.  The first
excited state is two--fold degenerate with an energy of 0.78~eV and is of the
same character as in silicon.

For the more massive proton the excited states are of lower energy. At the BC
site in silicon, the two--fold degenerate first excited state of the proton has
an energy of 0.27~eV, which is 0.07~eV higher than the ground state, while in
germanium the excited state has an energy of 0.25~eV, which is also 0.07~eV
above the ground state.

Each excited state of the muon/proton defines a different adiabatic potential
for the nuclei ({\it i.e.}, a different $E^{\mu}_{\alpha}({\bf
r}_{n})$ in Eq.~\ref{eq:H_I}).  It is therefore possible for
the lattice relaxations that occur when the muon/proton is in its first excited
state (say) to be different from those for the ground state.  For instance, the
fact that the wave function of the first excited state of the muon/proton is
essentially an excitation in the plane perpendicular to the bond, combined with
the fact that the potential well in this plane becomes narrower as the
separation of the NN atoms increases, results in the NN atoms actually relaxing
towards the impurity.  This relaxation is small for the muon, but effectively
zero for the proton due to the smaller zero--point energy.  The effect on the
energies of the excited states is negligible.

At the T site in silicon, the potential well is considerably shallower with a
depth of only 0.20~eV.  For the muon this means that even the ground state
energy (0.28~eV) of our constrained potential well (which is an upper bound on
the true ground state energy) is greater than the well depth. The proton,
however, has a (triply degenerate) first excited state with an energy of
0.14~eV which may therefore be bound at the T site.  In germanium the potential
well at the T site has a depth of just 0.18~eV in our 16--atom supercell
calculations. It follows from the behaviour found in going from the 16--atom to
the 54--atom supercell in silicon that the true well depth in germanium will
probably be less than this.  It is therefore unlikely that excited states of
either the muon or the proton will be bound at the T site in germanium.

\subsection{Energy barriers at the T and BC sites}
\label{barriers}

The heights of the energy barriers confining the muon and proton at the BC and
T sites are clearly of great importance in determining the dynamics of the
impurities within the lattice and hence are a significant part of the
configuration--coordinate diagram.

The static barriers ({\it i.e.}, excluding zero--point effects) experienced by
the muon and proton are identical.  For the BC site in silicon we calculate the
static barrier to motion towards the hexagonal site (in the [$-$1 1 0]
direction) to be 1.37~eV while in germanium it is 1.51~eV. The effective
barrier height is reduced by the zero--point energy and therefore depends on
the nature of the impurity.  Including this effect, the effective barrier
experienced by a muon at the BC site in silicon is 0.74~eV while in germanium
it is 0.95~eV.  For the proton, the effective barriers (1.17~eV in silicon and
1.33~eV in germanium) are higher. The effective barrier height for the muon at
the BC site may be considered a measure of the barrier to the BC$\rightarrow$T
site transition. In reality this transition is believed to involve charged
states: muonium at the BC site is first ionized (with activation energy
0.22~eV~\cite{Kreitzman95}) and then moves to the T site while simultaneously
recapturing an electron to regain its neutral charge state.  The sum of the
activation and barrier energies for these two processes as measured
experimentally is 0.60~eV.~\cite{Kreitzman95}

At the T site the energy barriers are very much lower.  In silicon the static
barrier to motion of the muon towards the hexagonal site (in the [1 1 1]
direction) is calculated to be 0.20~eV while in germanium it is 0.18~eV. When
zero--point effects are taken into account, the effective barriers for the muon
at the T site in silicon and germanium are zero indicating that, even at
$T=0$K, the muon is free to diffuse through the interstitial region.  However,
this barrier is not appropriate for the T$\rightarrow$BC site transition
because in our calculations for the muon in the interstitial region, the host
atoms around the BC site are unrelaxed. The process by which these atoms relax
the large distances required to allow the muon to move to the BC site is
unclear.  Experimentally, the barrier for the T$\rightarrow$BC site transition
in silicon is 0.39~eV.~\cite{Kreitzman95}

Since the zero--point energy of the proton is around a third of that of the
muon, these calculations indicate that it will be bound at the T site in both
silicon and germanium with an effective barrier of around 0.12~eV in each case.
As previously discussed the true effective barrier in germanium will probably
be lower than this.

\subsection{Hyperfine and superhyperfine parameters}
\label{hyperfine}

\begin{figure}
\caption{Spin density contour map in the neighbourhood of (a) the BC site and (b) the T site in silicon. In (a) the muon position is located dead centre, with the two nearest-neighbour silicon atoms above and below. In (b) the large concentration of positive spin density is located on the muon position, the positions of nearby silicon atoms in this plane are indicated with crosses. Continuous, dashed and dot-dashed lines correspond to positive, negative, and zero values respectively. The separation between adjacent isodensity contours is 0.001 e/bohr$^{3}$} 
\begin{center}
\epsfysize=9cm \epsfbox{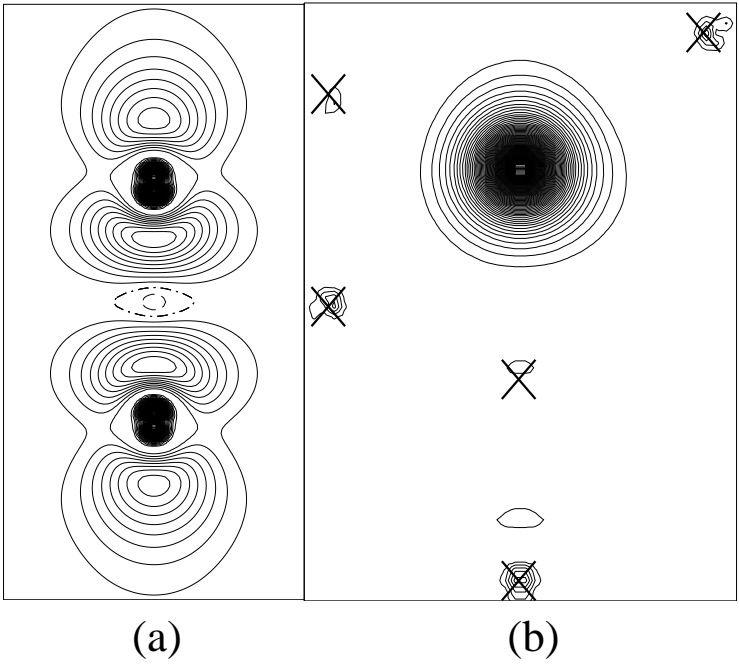}
\end{center}
\label{fig:spin_density}
\end{figure}

The hyperfine parameters depend on the spin density in the region at and around
the atomic nuclei. More specific insight into the origin of the large measured
differences between hyperfine parameters for muons located at the two impurity
sites can be gained from a consideration of the spin density isosurfaces.
Fig.~\ref{fig:spin_density} shows spin density contour plots for silicon in
appropriate planes encompassing the BC and T sites. Evidently the majority spin
density around an impurity placed at the BC site is largely dispersed onto the
two nearest--neighbour silicon atoms; the spin density in a small region around
the hydrogen nucleus is comparatively small and of opposite sign. At the T
site, by contrast, almost all of the majority spin density is localized on the
defect. From these calculations one therefore expects the isotropic hyperfine
parameter at the two sites to be of opposite sign, with the magnitude of the
parameter at the BC site much smaller than at the T site.

It is of course necessary to check the dependence of calculated hyperfine and
superhyperfine parameters on the supercell size and basis set quality. A set of
computed numbers are shown in Table~\ref{tble:hyp_convrg}. The parameters
appear to be reasonably well converged with respect to the basis set, but the
convergence with increasing supercell size is less good, particularly for the
isotropic hyperfine and superhyperfine parameters at the BC site.
Table~\ref{tble:hyp_BC} gives the hyperfine and superhyperfine parameters
calculated at the BC site in both silicon and germanium together with the
results of other calculations for comparison. Without motion averaging, the
values obtained for silicon using the LSDA approximation are in reasonable
agreement with both experiment and other DFT calculations.

\begin{figure}
\caption{Variation of the isotropic hyperfine parameter and the $xy$
component of ${\bf A}_{p}$ with displacement of the muon/proton from
the BC site.  The calculations were performed in silicon with the
16--atom supercell and the standard basis set.  The lines are
guides to the eye.} 
\begin{center}
\epsfysize=9cm \epsfbox{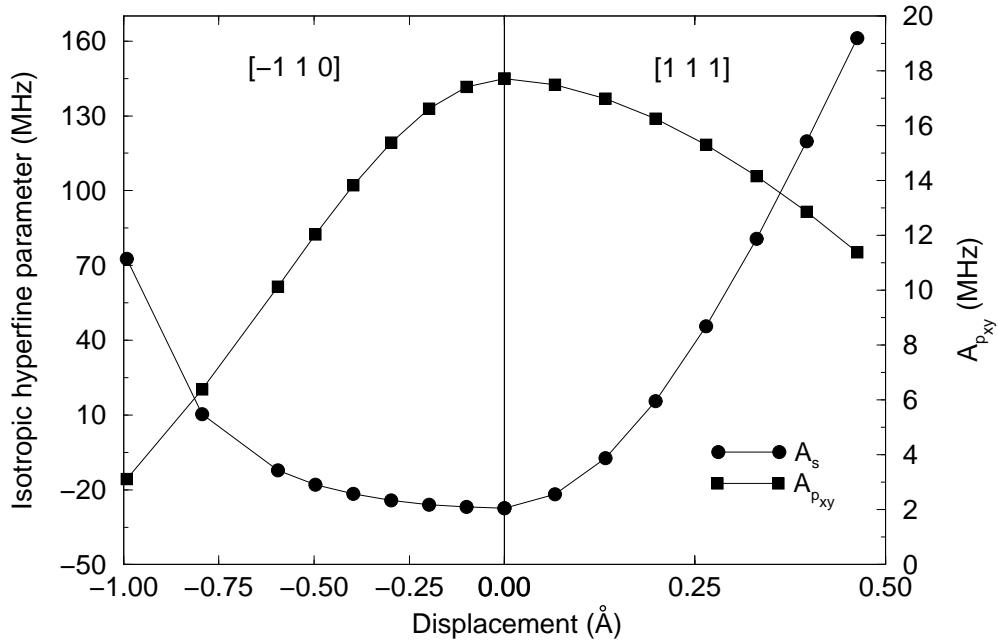}
\end{center}
\label{fig:hyper_bc}
\end{figure}

Both the hyperfine and superhyperfine motion--averaged tensors ($\langle {\bf
A}_{p_{\mu}}^{\rm BC} \rangle$ and $\langle {\bf A}_{p_{\rm Si}}^{\rm BC}
\rangle$) were found to be axially symmetric about the Si--Si bond ([1 1 1]
direction) in agreement with the experimental results.  As Luchsinger {\it et
al.}~\cite{Luch97} found, motion averaging increases the values of all but one
(the anisotropic hyperfine parameter) of the hyperfine and superhyperfine
parameters, with the isotropic (contact) term on the muon being the most
sensitive.  This is because of the very small contact charge density which
varies quite significantly with the muon position (Fig.~\ref{fig:hyper_bc}). In
agreement with Luchsinger {\it et al.}~\cite{Luch97}, use of the
Perdew--Wang~\cite{GGA} GGA functional did not consistently improve the values
of the parameters.  The results obtained for the muon at the BC site in
germanium follow a similar pattern.

The calculated hyperfine parameters for the T site are given in
Table~\ref{tble:hyp_T}. For silicon our values are in good agreement with
both experiment and previous calculations.  Again use of the
Perdew--Wang~\cite{GGA} GGA functional fails to improve this agreement.  For
germanium, our value of the isotropic hyperfine parameter at the T site also
agrees quite well with the measured value.

\begin{figure}
\caption{The variation in the isotropic
hyperfine parameter along the [1 1 1] direction at the T site in
silicon (54--atom supercell) and germanium (16--atom supercell). A positive displacement indicates movement towards the hexagonal site. The
lines are guides to the eye.}
\begin{center}
\epsfysize=9cm \epsfbox{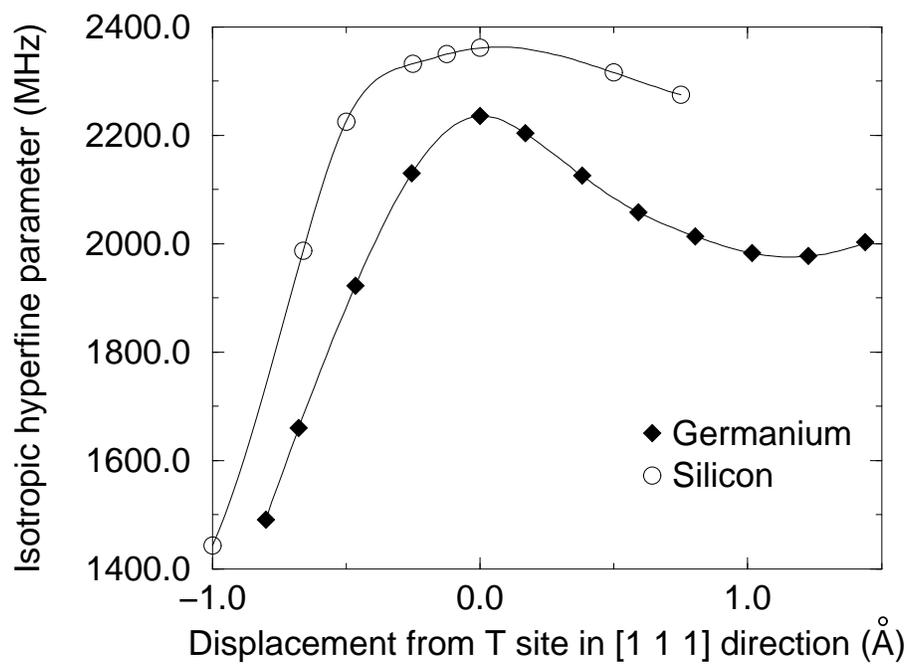}
\end{center}
\label{fig:hyper_t}
\end{figure}

The behaviour of the isotropic hyperfine parameter along the [1 1 1] axis in
the vicinity of the T site in silicon and germanium is shown in
Fig.~\ref{fig:hyper_t}. Motion averaging for the muon/proton at the T site
reduces the isotropic hyperfine parameter in both silicon and germanium.  The
final motion--averaged results are in reasonable agreement with experiment.
Motion averaging of ${\bf A}_{p_{\mu}}^{\rm T}$ resulted in an isotropic
tensor, in agreement with the symmetry arguments presented in
Section~\ref{hyperfine_mthd} and experimental observations.

In a recent application of the path--integral Monte Carlo approach, Miyake {\it
et al.}~\cite{Miyake_etal98} studied hydrogen and muonium at the T site in
silicon, with the electron--electron interactions calculated within the LDA.
They found the T site to be a local maximum in the potential energy surface, in
agreement with Luchsinger {\it et al.}~\cite{Luch97} but in disagreement with
our results and a recent plane--wave pseudopotential
calculation.~\cite{Probert} Their path--integral Monte Carlo study showed that
quantum effects led to the muonium distribution being centred on the T site
while hydrogen behaved as a largely classical particle and was thus distributed
away from the local maximum on that site. Evaluating the motion--averaged
isotropic hyperfine parameter with our hyperfine data gives a value of 492~MHz
for the hydrogen distribution of Miyake {\it et al.}, but 685~MHz with our
hydrogen distribution.  Therefore if one could measure the isotropic hyperfine
signal of {\it hydrogen} at the T site, one could deduce whether the T site is
a maximum or minimum in the potential energy surface.

\subsection{Energies of a muon/proton at the T and BC sites}
\label{energies}

The question of the relative stabilities of the muon and proton at the BC and T
sites is of considerable interest.  For a particular impurity this energy
difference is the sum of contributions from the static--lattice energy and the
zero--point energy.  The contribution from the static lattice
is sensitive to the size of the supercell and to the quality of the basis set. 
We investigated this point using the 16-- and 54--atom supercells.  We have
added a BSSE correction to each of the static--lattice energy differences
quoted here.  With the 16--atom silicon cell and the standard basis set, the T
site was found to be 0.63~eV lower in energy than the BC site.  Using the large
basis set reduced this to 0.32~eV.  In the 54--atom supercell and using the
standard basis set the T site was 0.41~eV lower in energy than the BC site. 
With the large basis set, this was reduced to just 0.07~eV.  These results
indicate that a 16--atom supercell is too small to give reliable estimates of
the static--lattice energy difference between the two sites. A summary of the
computed energies that influence the relative stabilities is given in
Table~\ref{tble:energy_summary}.

There have been several previous calculations of the static--lattice energy
difference between the T and BC sites in silicon.  Using a plane--wave
pseudopotential method and the LSDA, Chang and Chadi~\cite{chang89} found the T
site to be lower in energy, but only by an amount $\leq$0.25~eV.  Luchsinger
{\it et al.}~\cite{Luch97}, also using a plane--wave pseudopotential method,
found the T site to be 0.15~eV higher in energy than the BC site within the
LSDA and 0.19~eV higher within the GGA.  Note, however, that Luchsinger {\it et
al.}~\cite{Luch97} found the T site to be a local maximum in the energy and
that a nearby site has an energy about 0.05~eV lower.  It is clear from the
various results that the static--lattice energy difference between the T and BC
sites in silicon is small within the LSDA/GGA, but its precise value has yet to
be settled.

The fact that the static--lattice energy difference is small means that the
zero--point energy of the impurity is crucial in determining the relative
stability of the T and BC sites.  For a muon in silicon we have found the
zero--point energy at the BC site to be 0.35~eV larger than at the T site. 
This difference is large enough to to make the BC site unfavourable for the
muon, irrespective of which of the above values for the static--lattice energy
difference is used. However, the zero--point energy of the proton at the BC
site in silicon is only 0.12~eV higher than at the T site.  Therefore, for this
impurity the relative stability of the two sites depends on the precise value
of the static--lattice energy difference.

In germanium with a 16--atom supercell, the difference in static lattice
energies favours the T site by an energy of 0.57~eV. The convergence with
respect to supercell size found in silicon suggests that this difference in a
fully converged LSDA calculation would be smaller.  We estimate that the
zero--point energy of a muon at the BC site is 0.34~eV larger than at the T
site (where the form of the potential energy surface was obtained from data
calculated along the [1 1 1] axis only). For a proton the corresponding value
is 0.12~eV. These results are similar to those obtained in silicon and thus it
is likely that for the muon the T site is lower in energy. Without a fully
converged value for the static lattice energy difference we are unable to draw
any conclusions on the lowest energy site of the proton.

\section{Conclusions}
\label{conclusions}

We have calculated the zero--point motions and energies as well as the 
hyperfine parameters of both muonium and hydrogen when present as impurities in
silicon and germanium crystals at the BC and T sites.  The electron,
muon/proton and ion motions were decoupled using a double adiabatic
approximation, and for the BC site we have included the effect of the
zero--point motion on the relaxation of the host lattice. The ground states of
both the muon and proton at the BC sites of silicon and germanium are strongly
confined within a potential well of depth 1.37~eV (silicon) and 1.51~eV
(germanium). The zero--point energy of a muon at the BC site is calculated to
be 0.63~eV for silicon and 0.56~eV for germanium.  Despite the relatively large
zero--point energy of the muon at the BC site, it causes only a small
additional outwards relaxation of the nearest--neighbour silicon/germanium
atoms of about 0.01 \AA.  For the proton the additional relaxations of the
nearest neighbours due to zero--point motion are negligible.  At the T site the
static relaxations of the host atoms are very small and the zero--point energy
is considerably smaller than at the BC site, being 0.28~eV (0.22~eV) for a muon
in silicon (germanium).  It is therefore reasonable to assume that the
additional relaxation due to the zero--point motion is negligible for either a
muon or proton at the T site.

The relaxation of the crystal around either the BC or T sites is practically
independent of whether the impurity is a muon or a proton. This result confirms
one of the underlying assumptions of the widely accepted configuration
coordinate model.~\cite{config} The potential well at the BC sites of both
silicon and germanium is reasonably well described by a harmonic approximation,
at least for the ground states of the muon and proton.  The potential well at
the BC site in both materials is deep enough to bind several excited states of
the muon and proton, although we are not aware of any experimental evidence for
such states. The potential well at the T site in either silicon or germanium is
not deep enough to bind the muon which is free to diffuse through the
interstitial region, although our calculations suggest that the proton is bound
at this site at $T=0$~K. 

Various LSDA and GGA calculations have indicated that the energies for a static
muon or proton at the BC and T sites in silicon are very similar.  However, we
have calculated the difference in zero--point energies of a muon at the T and
BC sites in silicon (germanium) to be 0.35~eV (0.34~eV) which is sufficient to
make the T site more stable, whether we assume our value for the
static--lattice energy difference between the BC and T sites or those of
others.~\cite{chang89,Luch97} This result is in conflict with the
interpretation of experimental data.

The hyperfine parameters calculated for silicon in our all--electron
calculations are close to those obtained in plane--wave pseudopotential
calculations.  This agreement confirms that the procedure used to correct for
the pseudopotential and for the incomplete plane--wave basis sets are
accurate.  For silicon our static LSDA results are in reasonable agreement with
other LSDA results and experiment.  Our hyperfine parameter for the muon at the
T site in germanium is in much better agreement with experiment than the only
previous calculation of which we are aware.~\cite{casarin91}

In our work the motion averages of the hyperfine and superhyperfine parameters
are evaluated by averaging over the squared modulus of the wave function
obtained from the full solution of the muon/proton Schr\"odinger equation in
the potential well.  We note that the symmetry of the potential wells requires
that the symmetry of the motion--averaged hyperfine tensors at the T and BC
sites are the same as if the muon/proton was situated exactly at the sites.  We
have obtained detailed information about the variation of the hyperfine and
superhyperfine parameters with the position of the muon/proton.  Our results
show that motion averaging for the muon/proton at the BC site in silicon and
germanium increases the values of all of the hyperfine and superhyperfine
parameters apart from the anisotropic hyperfine term which decreases slightly,
in agreement with the conclusions of Luchsinger {\it et al}.~\cite{Luch97} 
With the exception of the isotropic hyperfine term however, all of the changes
are small. At the T sites in silicon and germanium, motion averaging reduces
the isotropic hyperfine parameter.

\section{Acknowledgments}

We thank R.~Q.~Hood and M.~I.~J.~Probert for useful discussions.
Financial support was provided by the Engineering and Physical
Sciences Research Council (UK).

\bibliographystyle{plain}

\begin{table}
\caption{The values of the parameters in Eq.~\ref{eq:BCexpn} defining
the potential well at the BC site of silicon and germanium.  The fit
is applicable within a cylinder centred on the BC site of radius 1.0~\AA \
(1.1~\AA) in the plane perpendicular to the bond and up to
$\pm$0.5~\AA \ (0.37~\AA) along the direction of the bond in silicon
(germanium).  The units are such that if the lengths in
Eq.~\ref{eq:BCexpn} are expressed in Bohr radii then the potential
energy is in Ha. N.B. The values quoted are those carried into the zero--point calculation and therefore the number of significant figures should not be taken as an indication of the accuracy of the fit.}
\label{tble:BC_param}
\begin{tabular}{ldd}
& Silicon & Germanium \\
\tableline
$\beta$  & 0.00588753  & 0.00610711 \\
$\gamma$ & 0.0807689  & 0.0548517 \\
$\delta$ & 0.0        & 0.0   \\
$\zeta$  & 0.000337036  & 0.000224779 \\
$\eta$   & 0.0654643  & 0.0506238 \\
\end{tabular}
\end{table}

\begin{table}
\caption{The parameters defining the expansions (Eq.~\ref{eq:Texpn}) of
the potential energy surface around the T site in silicon and
germanium.  The fit is applicable over the region bounded by a sphere
of radius 1.0~\AA\ centred on the T site.  The units are such that if
the lengths in Eq.~\ref{eq:Texpn} are expressed in Bohr radii then the
potential energy is in Ha. N.B. The values quoted are those carried into the zero--point calculation and therefore the number of significant figures should not be taken as an indication of the accuracy of the fit.}
\label{tble:T_param}
\begin{tabular}{ldd}
         & Silicon       & Germanium   \\
\tableline
$a_{2}$  & 0.00440492     & 0.00151269   \\ 
$a_{3}$  & 0.00589898     & 0.00542489   \\ 
$a_{4}$  & $-$0.00197527  & 0.000339949  \\ 
$b_{4}$  & 0.000930883    & $-$0.000160207 \\ 
$a_{5}$  & $-$0.00456177  & $-$0.00311144  \\ 
$a_{6}$  & 0.00155753     & 0.000775975  \\ 
$b_{6}$  & $-$0.000556122 & $-$0.000277064 \\ 
$c_{6}$  & 0.00696501     & 0.00347002   \\ 
\end{tabular}
\end{table}
\clearpage
\mediumtext
\begin{table}
\caption{The dependence of hyperfine and superhyperfine parameters for
muonium in silicon on supercell size and basis set.}  
\label{tble:hyp_convrg}
\begin{tabular}{llddddd}
 & & \multicolumn{4}{c}{BC site} & T site \\
Supercell & Basis set & $A_{s_{\mu}}$ & $A_{p_{\mu}}$ & $A_{s_{\mathrm Si}}$ & 
$A_{p_{\mathrm Si}}$ & $A_{s_{\mu}}$ \\
\tableline
16 atom    & Standard     & $-$27.1 & 17.7 & $-$147. & $-$13.6 & 2302 \\
16 atom    & Large & $-$21.4 & 13.0 & $-$114. & $-$10.7 & 2366 \\
54 atom    & Standard     & $-$1.6 & 16.4 & $-$91.0 & $-$12.4 & 2362 \\
54 atom    & Large & 4.5    & 9.8 & $-$57.1 & $-$8.1 & 2389 \\
\end{tabular}
\end{table}

\begin{table}
\caption{Static and motion--averaged (indicated by
$\langle\rangle$) hyperfine parameters for the muon at the BC site and
the nearest--neighbour atoms.  PS denotes a pseudopotential
calculation.}
\label{tble:hyp_BC}
\begin{tabular}{ldddddddd}
 & \multicolumn{8}{c}{Hyperfine parameters (MHz)} \\
 & \multicolumn{4}{c}{Silicon} & \multicolumn{4}{c}{Germanium} \\
 & $A_{s_{\mu}}$ & $A_{p_{\mu}}$ & $A_{s_{\mathrm Si}}$ & 
$A_{p_{\mathrm Si}}$ & $A_{s_{\mu}}$ & $A_{p_{\mu}}$ & 
$A_{s_{\mathrm Ge}}$ & $A_{p_{\mathrm Ge}}$  \\
\tableline
LSDA\tablenotemark[1]             & $-$1.6 & 16.4 & $-$91.0 &
$-$12.4  &  &  &  &  \\
LSDA\tablenotemark[2]             & $-$27.1 & 17.7 & $-$147. &
$-$13.6  & $-$24.6 & 16.4 & $-$80.7 & $-$5.8 \\
$\langle$LSDA$\rangle$\tablenotemark[2]
& 2.5  & 14.5 & $-$141. & $-$13.4  & 3.6 & 12.5 & $-$75.5 & $-$5.6 \\
GGA\tablenotemark[2]  
& $-$89.3 & 18.9 & $-$155. & $-$14.0 & $-$64.6 & 17.1 & $-$81.0 & 
$-$5.9\\
LSDA\tablenotemark[3]
& $-$104.  & 58.5  &  $-$127. & $-$53.5 & $-$87. & 64. & $-$85. & $-$24.\\
PS-LSDA\tablenotemark[4]   
& $-$26.   & 22.8 &  $-$90.  & $-$20.2    & & & &\\
PS-GGA\tablenotemark[4]
& $-$81.   & 27.5 &  $-$192. & $-$28.      & & & &\\
$\langle$PS-GGA$\rangle$\tablenotemark[4]
& $-$65. & 21.7 &  $-$191. & $-$26.2       & & & &\\
PS-LSDA\tablenotemark[5] 
& $-$26.8  & 18.1  &  $-$83.8 & $-$22.7 & & & &\\
PS-LSDA\tablenotemark[6]
& $-$35.  & 22.3  &  $-$85. & $-$21.5     & & & &\\
Experiment\tablenotemark[7]
& $-$67.3 & 25.3 & $-$95.1 & $-$21.2    & & & &\\ 
Experiment\tablenotemark[8]
& & &  &  & $-$96.5 & 34.6 & &\\ 
\end{tabular}
\tablenotemark[1]{This work, 54--atom supercell and ``standard'' basis set.} \\
\tablenotemark[2]{This work, 16--atom supercell and ``standard'' basis set.} \\
\tablenotemark[3]{Casarin {\it et al.}~\cite{casarin91}} \\
\tablenotemark[4]{Luchsinger {\it et al.}~\cite{Luch97}} \\
\tablenotemark[5]{Van de Walle.~\cite{VdW90}} \\ 
\tablenotemark[6]{Van de Walle and Bl\"{o}chl.~\protect\cite{VdW_Blochl93}} \\
\tablenotemark[7]{Kiefl and Estle.~\protect\cite{Kiefl}} \\
\tablenotemark[8]{Patterson.~\protect\cite{Pat88}} \\
\end{table}
\clearpage
\begin{table}
\caption{Isotropic hyperfine parameters for the muon at the T site of
silicon and germanium.  The quoted results are for the ``standard''
basis set.}
\label{tble:hyp_T}
\begin{tabular}{lcc}
& \multicolumn{2}{c}{$A_{s_{\mu}}$ (MHz)} \\
& Silicon & Germanium \\
\tableline
LSDA\tablenotemark[1]       & 2302 & 2236 \\
GGA\tablenotemark[1]        & 2651 & 2548 \\
$\langle{\mathrm LSDA}\rangle$\tablenotemark[1] & 2096 & 2032\\
LSDA\tablenotemark[2]        & 2362 &      \\
$\langle{\mathrm LSDA}\rangle$\tablenotemark[2] & 2152 & \\
PS--LSDA\tablenotemark[3]    & 1939 &      \\
PS--GGA\tablenotemark[3]     & 2098 &      \\
LSD--VBH\tablenotemark[4]    & 3043 & 3977 \\
Experiment\tablenotemark[5] & 2006 & 2360 \\ 
\end{tabular}
\tablenotemark[1]{This work, 16--atom supercell.} \\
\tablenotemark[2]{This work, 54--atom supercell.} \\
\tablenotemark[3]{Luchsinger \textit{et al.}~\protect\cite{Luch97}}
\\
\tablenotemark[4]{Casarin {\it et al.}~\cite{casarin91}, extended
basis set.} \\
\tablenotemark[5]{Patterson~\protect\cite{Pat88} and references therein.} \\
\end{table}

\begin{table}
\caption{A summary of the energies influencing the relative
stability of the BC and T sites in silicon and germanium.  The
importance of the zero--point energy of the muon in determining the favoured site is clear.}
\label{tble:energy_summary}
\begin{tabular}{lcccc}
 & \multicolumn{2}{c}{Silicon} & \multicolumn{2}{c}{Germanium} \\
 & BC site & T site & BC site & T site \\
\tableline
Static lattice energy w.r.t. BC site (eV) & 0   &
$-$0.07\tablenotemark[1] & 0 & $-$0.57\tablenotemark[2] \\
Muon zero--point energy (eV) & 0.63\tablenotemark[3] 
& 0.28\tablenotemark[4] & 0.56\tablenotemark[2] &
0.22\tablenotemark[2] \\
Total energy w.r.t. BC site for muon (eV) & 0 & $-$0.42 & 0 &
$-$0.91 \\
Proton zero--point energy (eV) & 0.20\tablenotemark[3] &
0.09\tablenotemark[4]  & 0.18\tablenotemark[2] & 0.06\tablenotemark[2]
\\
Total energy w.r.t. BC site for proton (eV) & 0 & $-$0.18 & 0 &
$-$0.69 \\
\end{tabular}
\tablenotemark[1]{54--atom supercell and ``large'' basis set.} \\
\tablenotemark[2]{16--atom supercell.} \\
\tablenotemark[3]{16--atom supercell and ``standard'' basis set.} \\
\tablenotemark[4]{54--atom supercell and ``standard'' basis set.} \\
\end{table}
\clearpage

\clearpage

\clearpage

\clearpage

\clearpage

\clearpage

\end{document}